\documentstyle[aps,prl,epsfig,12pt]{revtex}

\begin{document}

\epsfysize3cm
\epsfbox{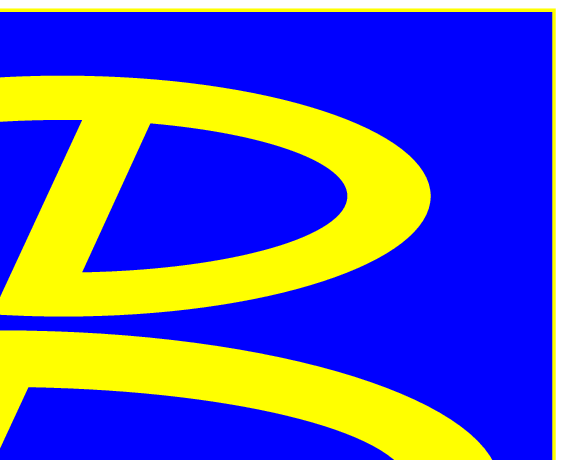}    

\vskip -3cm
\noindent
\hspace*{12cm}KEK Preprint 2001-11 \\
\hspace*{12cm}Belle Preprint 2001-5 \\

\vskip 1cm

\renewcommand{\baselinestretch}{1.2}

\begin{center}

{\Large\bf
	Measurement of Branching Fractions for\\
	$B\rightarrow \pi\pi$, $K\pi$ and $KK$ Decays
	\footnote{submitted to PRL}
}

\vskip 0.5cm

(The Belle Collaboration)\\

\vskip 0.2cm

{\normalsize
K.~Abe$^{10}$, 
K.~Abe$^{37}$, 
I.~Adachi$^{10}$, 
Byoung~Sup~Ahn$^{15}$, 
H.~Aihara$^{38}$, 
M.~Akatsu$^{20}$, 
G.~Alimonti$^{9}$, 
Y.~Asano$^{42}$, 
T.~Aso$^{41}$, 
V.~Aulchenko$^{2}$

% ...
$^{2}${Budker Institute of Nuclear Physics, Novosibirsk}\\
% ...
$^{9}${University of Hawaii, Honolulu HI}\\
$^{10}${High Energy Accelerator Research Organization (KEK), Tsukuba}\\
% ...
$^{15}${Korea University, Seoul}\\
% ...
$^{20}${Nagoya University, Nagoya}\\
% ...
$^{37}${Tohoku University, Sendai}\\
$^{38}${University of Tokyo, Tokyo}\\
% ...
$^{41}${Toyama National College of Maritime Technology, Toyama}\\
$^{42}${University of Tsukuba, Tsukuba}\\
% ...
}

\end{center}

\centerline{Abstract}
\vskip -0.5cm
\begin{abstract}

We report measurements of the branching fractions for
$B^0\rightarrow\pi^+\pi^-$, $K^+\pi^-$, $K^+K^-$ and $K^0\pi^0$, and
$B^+\rightarrow\pi^+\pi^0$, $K^+\pi^0$, $K^0\pi^+$ and $K^+\overline{K}{}^0$.
The results are based on 10.4 fb$^{-1}$ of data collected on the
$\Upsilon$(4S) resonance at the KEKB $e^+e^-$ storage ring with the
Belle detector, equipped with a high momentum particle identification
system for clear separation of charged $\pi$ and $K$ mesons.
We find
${\cal B}(B^0\rightarrow\pi^+\pi^-)
=(0.56^{\ +0.23}_{\ -0.20}\pm 0.04)\times 10^{-5}$,
${\cal B}(B^0\rightarrow K^+\pi^-)
=(1.93^{\ +0.34\ +0.15}_{\ -0.32\ -0.06})\times 10^{-5}$,
${\cal B}(B^+\rightarrow K^+\pi^0)
=(1.63^{\ +0.35\ +0.16}_{\ -0.33\ -0.18})\times 10^{-5}$,
${\cal B}(B^+\rightarrow K^0\pi^+)
=(1.37^{\ +0.57\ +0.19}_{\ -0.48\ -0.18})\times 10^{-5}$, and
${\cal B}(B^0\rightarrow K^0\pi^0)
=(1.60^{\ +0.72\ +0.25}_{\ -0.59\ -0.27})\times 10^{-5}$,
where the first and second errors are statistical and systematic.
We also set upper limits of
${\cal B}(B^+\rightarrow\pi^+\pi^0)<1.34\times 10^{-5}$,
${\cal B}(B^0\rightarrow K^+K^-)<0.27\times 10^{-5}$, and
${\cal B}(B^+\rightarrow K^+\overline{K}{}^0)<0.50\times 10^{-5}$
at the $90\%$ confidence level.

\vspace{0.5cm}

\end{abstract}

\normalsize


The charmless hadronic $B$ decays $B\rightarrow\pi\pi$, $K\pi$ and
$KK$ provide a rich sample to test the standard model and to probe
new physics \cite{phys}.
Of particular interest are indirect and direct $CP$ violation in
the $\pi\pi$ and $K\pi$ modes, which are related to the angles
$\phi_2$ and $\phi_3$ of the unitarity triangle, respectively \cite{phys}.
Measurements of branching fractions of these decay modes are an
important first step toward these $CP$ violation studies.
However, experimental information is rather limited, and the only
published results come from one experiment \cite{cleo}.
One of the key experimental issues is the particle identification (PID)
for separation of the high momentum charged $\pi$ and $K$ mesons.
This is one of the primary reasons that the $B$ factory experiments
\cite{ichep_belle,ichep_babar} have been equipped with specialized high
momentum PID devices.

In this paper, we report the first results of the Belle experiment on
charmless hadronic two-body $B$ decays into $\pi\pi$, $K\pi$ and $KK$
final states.
The decay modes studied are $\pi^+\pi^-$, $K^+\pi^-$, $K^+K^-$
and $K^0\pi^0$ for $B^0$ decays, and $\pi^+\pi^0$, $K^+\pi^0$, $K^0\pi^+$,
$K^+\overline{K}{}^0$, for $B^+$ decays.
For the modes with $K^0$ mesons, only $K^0_S\rightarrow\pi^+\pi^-$ decays
are used.
Throughout this paper, the inclusion of charge conjugate states is implied.
The results are based on data taken by the Belle detector \cite{belle}
at the KEKB asymmetric $e^+e^-$ storage ring \cite{kekb}.
The Belle detector is equipped with aerogel \v{C}erenkov counters (ACC)
configured for high momentum PID.
The data set consists of 10.4 fb$^{-1}$ data taken at the $\Upsilon$(4S)
resonance, corresponding to 11.1 million $B\overline{B}$ events,
and 0.6 fb$^{-1}$ data taken at an energy $\sim$60 MeV below the resonance,
for systematic studies of the continuum $q\overline{q}$ background.

Primary charged tracks are required to satisfy track quality cuts
based on their impact parameters relative to the interaction point (IP).
$K^0_S$ mesons are reconstructed using pairs of charged tracks
that have an invariant mass within $\pm$30 MeV/$c^2$ of the known
$K^0_S$ mass and a well reconstructed vertex that is displaced from the IP.
Candidate $\pi^0$ mesons are reconstructed using $\gamma$ pairs with an
invariant mass within $\pm$16 MeV/$c^2$ of the nominal $\pi^0$ mass.
The $B$ meson candidates are reconstructed using
the beam constrained mass, $m_{bc}=\sqrt{E_{\rm beam}^2-p_B^2}$,
and the energy difference, $\Delta E=E_B-E_{\rm beam}$, where
$E_{\rm beam}\equiv\sqrt{s}/2\simeq 5.290$ GeV, and
$p_B$ and $E_B$ are the momentum and energy of the reconstructed $B$
in the $\Upsilon$(4S) rest frame, respectively.
The signal region for each variable is defined as $\pm 3\sigma$ from
its central value.
The resolution in $m_{bc}$ is dominated by the beam energy spread and
is typically 2.7 MeV/$c^2$.
The $\Delta E$ resolution ranges from 20 to 25 MeV, depending on
the momentum and energy resolutions for each particle.
Normally we compute $\Delta E$ assuming a $\pi$ mass for each charged
particle.
This shifts $\Delta E$ downward by 44 MeV for each charged $K$ meson,
giving kinematic separation between the $h\pi^+$ and $hK^+$ $(h=\pi, K)$
final states.
In modes with $\pi^0$ mesons, both the $m_{bc}$ and $\Delta E$ distributions
are asymmetric due to $\gamma$ interactions in the material in front of the
calorimeter and energy leakage out of the calorimeter.
We accept events in the region $m_{bc}>5.2$ GeV/$c^2$ and
$|\Delta E|<0.25$ GeV for the $h^+h^-$ and
$K^0_Sh^+$ modes, and $-0.45<\Delta
E<0.15$ GeV for the $h^+\pi^0$ and $K^0_S\pi^0$ modes.
In this kinematic window, the area outside the signal region is defined
as a sideband.
The signal reconstruction efficiencies after the kinematic window cut
are 65\% for $h^+h^-$, 33\% for $K^0_Sh^+$, 50\% for $h^+\pi^0$, and
24\% for $K^0_S\pi^0$, according to a GEANT \cite{geant} based Monte Carlo
(MC) simulation.
The MC tracking efficiency is verified by detailed studies using high
momentum tracks from $D$, $\eta$ and $K^*$ decays.
The reconstruction efficiencies for high momentum $K^0_S$ and $\pi^0$
mesons are tested by comparing the ratio of the yield of
$D^+\rightarrow K^0_S\pi^+$ to $D^+\rightarrow K^-\pi^+\pi^+$ and
$D^0\rightarrow K^-\pi^+\pi^0$ to $D^0\rightarrow K^-\pi^+$,
respectively, between data and MC simulation.
From these studies, we assign a relative systematic error in these
efficiencies of 2.3\% per charged track, 12\% per $K^0_S$ and 8.5\%
per $\pi^0$ meson.

The background from $b\rightarrow c$ transitions is negligible.
The dominant background is from the continuum $q\overline{q}$ process.
We suppress this background using the event topology, which is spherical for
$B\overline{B}$ events and jet-like for $q\overline{q}$ events in the
$\Upsilon$(4S) rest frame.
This difference can be quantified by using several variables including the
event sphericity, $S$, the angle between the $B$ candidate thrust axis and
the thrust axis of the rest of the event, $\theta_T$, and the Fox-Wolfram
moments \cite{fw}
$H_l=\sum_{i,j}{|\vec{p_i}||\vec{p_j}|P_l(\cos\theta_{ij})}$,
where the indices $i$ and $j$ run over all final state particles,
$\vec{p}_i$ and $\vec{p}_j$ are the momentum vectors of particles $i$
and $j$, $P_l$ is the $l$-th Legendre polynomial, and $\theta_{ij}$ is the
angle between particles $i$ and $j$.
We can also use the $B$ flight direction, $\theta_B$, and the decay axis
direction, $\theta_{hh}$, which distinguish $B\overline{B}$ from
$q\overline{q}$ processes based on initial state angular momentum.

We increase the suppression power of the normalized Fox-Wolfram moments,
$R_l=H_l/H_0$, by decomposing them into three terms:
$R_l=R_l^{ss}+R_l^{so}+R_l^{oo}=(H_l^{ss}+H_l^{so}+H_l^{oo})/H_0$,
where the indices $ss$, $so$, and $oo$ indicate respectively that
both, one, or neither of the particles comes from a $B$ candidate.
These are combined into a six term Fisher discriminant \cite{fisher}
called the Super Fox-Wolfram \cite{sfw} defined as
$SFW=\sum^{4}_{l=1}{(\alpha_lR_l^{so}+\beta_lR_l^{oo})}$,
where $\alpha_l$ and $\beta_l$ are Fisher coefficients and $l$=2,4 for
$\alpha_l$ and $R_l^{so}$.
The terms $R_l^{ss}$ and $R_{l=1,3}^{so}$ are excluded because they are
strongly correlated with $m_{bc}$ and $\Delta E$.
In the $h^+h^-$ modes, for example, $SFW$ gives a 20\% increase in
the expected significance compared to $R_2$.

We combine different $q\overline{q}$ suppression variables into a single
likelihood,
${\cal L}_{s(q\overline{q})}=\prod_i{{\cal L}_{s(q\overline{q})}^i}$,
where the ${\cal L}_{s(q\overline{q})}^i$ denotes the signal($q\overline{q}$)
likelihood of the suppression variable $i$, and select candidate events
by cutting on the likelihood ratio
${\cal R}_s={\cal L}_s/({\cal L}_s+{\cal L}_{q\overline{q}})$.
For $h^+h^-$ and $K^0_Sh^+$, the likelihood contains $SFW$,
$\cos\theta_B$, and $\cos\theta_{hh}$.
In modes with $\pi^0$ mesons, the $q\overline{q}$ background is
significantly larger.
In this case, we first make a loose cut on $\cos\theta_T$.
Next, we extend $SFW$ to include $\cos\theta_T$ and $S$, and form the
likelihood using this extended $SFW$ and $\cos\theta_B$.
In each case, the signal probability density functions (PDFs) are
determined using MC simulation and the $q\overline{q}$ PDFs are taken
from $m_{bc}$ sideband data.
The performance of ${\cal R}_s$ varies among the modes with
efficiencies ranging from $40$\% to $51\%$ while removing more than $95\%$
of the $q\overline{q}$ background.
The $\pi^+\pi^0$ mode calls for a tighter cut with an efficiency of $26\%$.
The error in these efficiencies is determined by applying the same
procedure to the $B^+\rightarrow\overline{D}{}^0\pi^+$,
$\overline{D}{}^0\rightarrow K^-\pi^+$ event sample and comparing the
cut efficiencies between data and MC.
The relative systematic error is determined to be $4\%$.

The high momentum charged $\pi$ and $K$ mesons
($1.5<p_{h^{\pm}}<4.5$ GeV/$c$ in the laboratory frame)
are distinguished by cutting on the $\pi (K)$ likelihood ratio
${\cal R}_{\pi(K)}\equiv{\cal L}_{\pi (K)}/({\cal L}_{\pi}+{\cal L}_K)$,
where ${\cal L}_{\pi (K)}$ denotes the product of each $\pi (K)$ likelihood
of their energy loss ($dE/dx$) in the central drift chamber 
and their \v{C}erenkov light yield in the ACC.
Each likelihood is calculated from a PDF determined
using MC simulation.
The PID efficiency and fake rate are measured using $\pi$ and $K$
tracks in the same kinematic range as signal,
with kinematically selected $D^{*+}\rightarrow D^0\pi^+$,
$D^0\rightarrow K^-\pi^+$ decays.
The efficiency and fake rate for $\pi$ mesons are measured to be 92\%
and 4\% (true $\pi$ fakes $K$), whereas those for $K$ mesons are 85\%
and 10\% (true $K$ fakes $\pi$), respectively.
The relative systematic error in the PID efficiency is 2.5\%
per charged $\pi$ or $K$ meson.

\begin{figure}[]
  \begin{center}
  \epsfxsize 8.5cm
  \epsfbox{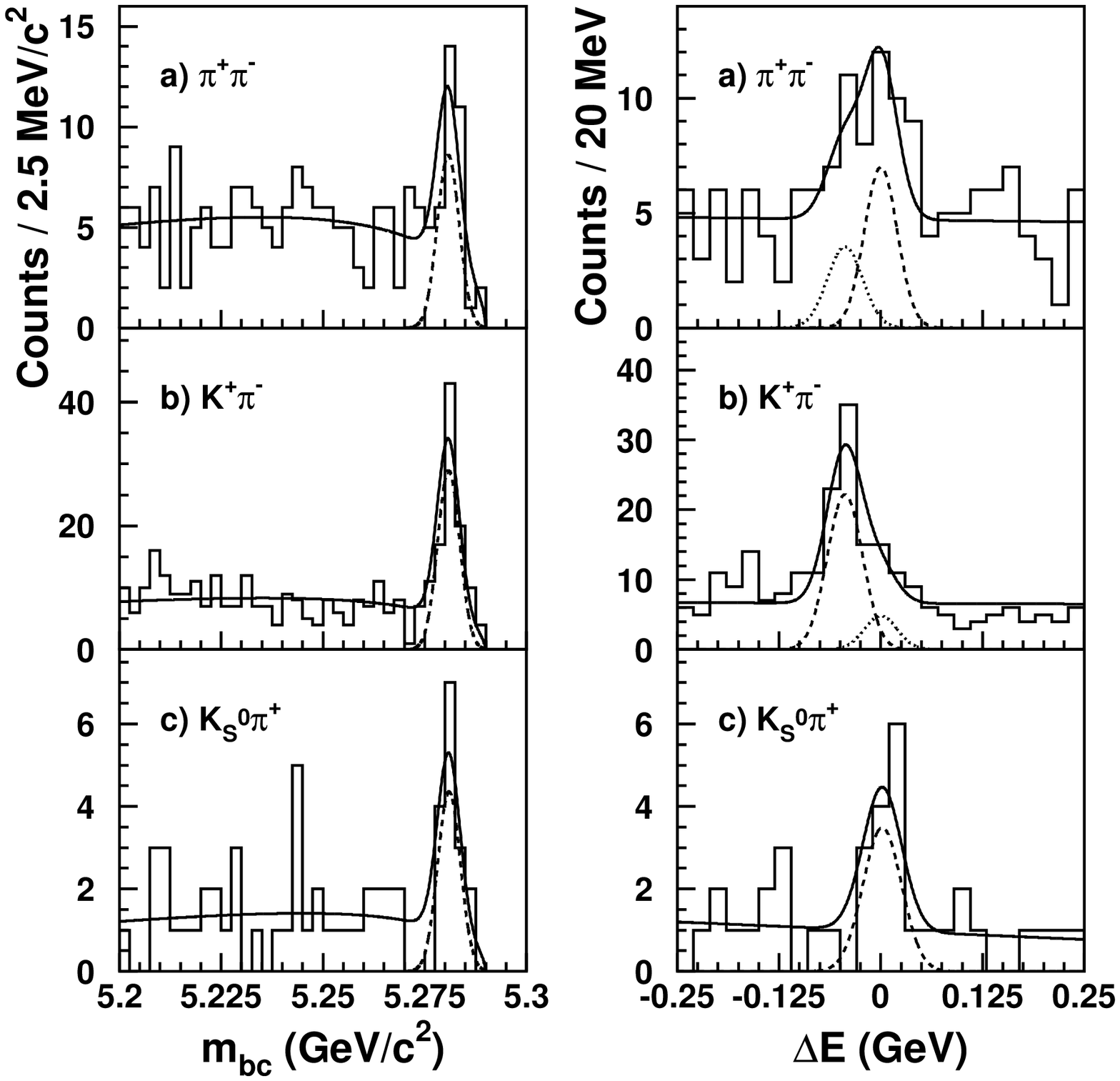}
  \end{center}
  \caption{{}
	The $m_{bc}$ (left) and $\Delta E$ (right) distributions,
	in the signal region of the other variable, for $B\rightarrow$
	a) $\pi^+\pi^-$, b) $K^+\pi^-$ and c) $K^0_S\pi^+$.
	The fit function and its signal component are shown by the solid
	and dashed curve, respectively. 
	In the $\pi^+\pi^-$ and $K^+\pi^-$ fits, the cross-talk components
	are shown by dotted curves.
  }
\label{fig:fig1}
\end{figure}

Figure \ref{fig:fig1} shows the $m_{bc}$ and $\Delta E$ distributions
in the signal region of the other variable, for the $\pi^+\pi^-$,
$K^+\pi^-$ and $K^0_S\pi^+$ modes.
Each $m_{bc}$ and $\Delta E$ distribution is fitted to a Gaussian signal
plus a background function.
The $m_{bc}$ and $\Delta E$ peak positions and $m_{bc}$ width are
calibrated using the
$B^+\rightarrow\overline{D}{}^0\pi^+$,
$\overline{D}{}^0\rightarrow K^+\pi^-$ data sample.
The $\Delta E$ Gaussian width is calibrated using high momentum
$D^0\rightarrow K^-\pi^+$ and
$D^+\rightarrow K^0_S\pi^+$ decays.
The $m_{bc}$ background shape is modeled by the ARGUS background function
\cite{argus} with parameters determined using positive $\Delta E$ sideband
data.
A linear function is used to model the shape of the $\Delta E$ background;
the slope is fixed at the value determined from the $m_{bc}$ sideband.
The signal yields are determined from the $\Delta E$ fits where there is
kinematic separation between the $h\pi^+$ and $hK^+$ decays.
The $\pi^+\pi^-$ and $K^+\pi^-$ fits include a component to account
for misidentified backgrounds.  
The normalizations of these components are free parameters.
The extracted yields are listed in Table \ref{tab:table1}.
The cross-talk among different signal modes is consistent with expectations
based on PID fake rates.
No excess is observed in the $K^+K^-$ and $K^+K^0_S$ modes.

\begin{figure}[]
  \begin{center}
  \epsfxsize 8.5cm
  \epsfbox{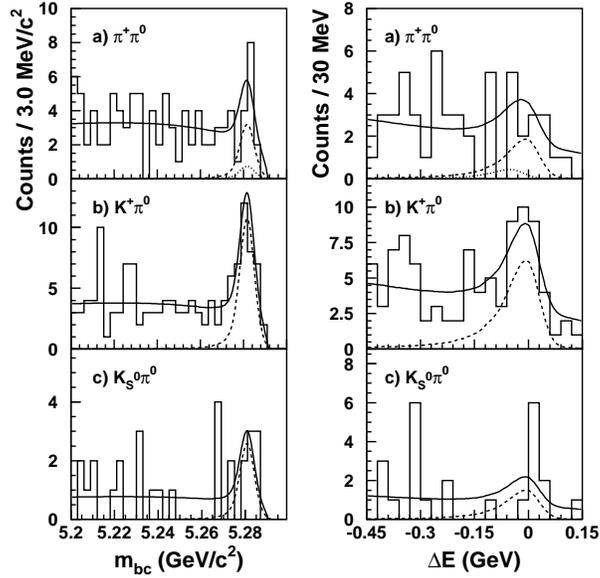}
  \end{center}
\caption{{}
	The $m_{bc}$ (left) and $\Delta E$ (right) projections for
	$B\rightarrow$ a) $\pi^+\pi^0$, b) $K^+\pi^0$ and c) $K^0_S\pi^0$.
	For $K^+\pi^0$, a $K$ mass is assumed for the charged particle.
	The projection of the two-dimensional fit onto each variable and
	its signal component are shown by the solid and dashed curve,
	respectively. 
	In the $\pi^+\pi^0$ fit, the cross-talk from $K^+\pi^0$
	is indicated by a dotted curve.
}  
\label{fig:fig2}
\end{figure}

Figure \ref{fig:fig2} shows the $m_{bc}$ and $\Delta E$ projections
for the $\pi^+\pi^0$, $K^+\pi^0$ and $K^0_S\pi^0$ modes.
For these modes, since the $\Delta E$ distribution has a long tail,
a two-dimensional fit is applied to the $m_{bc}$ and $\Delta E$
distributions.
The signal distribution is modeled by a smoothed two-dimensional MC
histogram, while the background distribution is taken to be the product
of the $m_{bc}$ and $\Delta E$ background functions discussed above.
The signal and background shapes are determined following the same procedure
as for the $h^+h^-$ and $K^0_Sh^+$ modes.
The $\Delta E$ resolution is calibrated using $D^0\rightarrow K^-\pi^+\pi^0$
decays where the $\pi^0$ is reconstructed in the same kinematic range as
the signal.
For the $\pi^+\pi^0$ mode, since the cross-talk from $K^+\pi^0$ is expected
to be large and the $\Delta E$ separation is less than 1$\sigma$, the
$K^+\pi^0$ component is fixed at its expected level.
The obtained yields are listed in Table \ref{tab:table1}.

The systematic error in the signal yield is determined by varying the
parameters of the fitting functions within $\pm 1\sigma$ of their nominal
values.
The changes in the signal yield from each variation are added in quadrature.
These errors range from 1\% to 6\%.
In the $K^+\pi^-$ mode, the $\Delta E$ background normalization is
influenced by an excess around $-175$ MeV.
In this region, we expect to observe a few background events from
$B$ decays such as $B\rightarrow\rho\pi$, $K^*\pi$, and $K^*\gamma$
(for modes with $\pi^0$ mesons),
based on a MC simulation \cite{pdg,ichep_cleo} in all signal modes.
To estimate their effect, we either exclude the negative $\Delta E$
sideband from the fit or add these components to the fit based on
MC histograms.
The resulting change in the signal yield, ranging from 4\% to 10\%,
is added in quadrature to the above systematic error.

Table \ref{tab:table1} summarizes all results.
The statistical significance ($\Sigma$) is defined as
$\sqrt{-2\ln ({\cal L}(0)/{\cal L}_{\rm max})}$, where
${\cal L}_{\rm max}$ and ${\cal L}(0)$ denote the maximum likelihood
with the nominal signal yield and with the signal yield fixed at zero,
respectively \cite{pdg}.
The final systematic error is the quadratic sum of the relative error in
the signal yield ($N_s$), the reconstruction, PID, and continuum suppression
efficiencies, and the number of $B\overline{B}$ pairs (1\%). 
If $\Sigma <3$, we set a $90\%$ confidence level upper limit on the signal
yield ($N_s^{\rm U.L.}$) from the relation
$\int^{N_s^{\rm U.L.}}_0{{\cal L}(N_s)dN_s}/
\int^{\infty}_0{{\cal L}(N_s)dN_s}=0.9$, where ${\cal L}(N_s)$ denotes
the maximum likelihood with the signal yield fixed at $N_s$.
The branching fraction upper limit (U.L.) is then calculated by increasing
$N_s^{\rm U.L.}$ and reducing the efficiency by their systematic errors.

In summary, using 11.1 million $B\overline{B}$ events recorded in the
Belle detector, the charge averaged branching fractions for
$B\rightarrow \pi^+\pi^-$, $K^+\pi^-$, $K^+\pi^0$, $K^0\pi^+$,
and $K^0\pi^0$ are measured with statistically significant signals.
For the $\pi^+\pi^0$ mode, an excess is seen with marginal significance.
No excess is observed for the $K^+K^-$ and $K^+\overline{K}{}^0$ modes.
For these modes, 90\% confidence level upper limits are set.
The results are listed in Table \ref{tab:table1}.
In Table \ref{tab:table2}, we list some ratios of branching fractions
based on these measurements.
Recent theoretical work \cite{phys} suggests that the ratio
${\cal B}(B^+\rightarrow \pi^+\pi^0)/{\cal B}(B^0\rightarrow \pi^+\pi^-)$
is relevant for extracting $\phi_2$, the ratio
${\cal B}(B^+\rightarrow K^+\pi^0)/{\cal B}(B^0\rightarrow K^+\pi^-)$
is relevant for determining the contribution from electro-weak penguins,
and the remaining four ratios are useful to constrain $\phi_3$.
All the branching fraction and ratio results are consistent with other
measurements \cite{cleo,ichep_babar}.
Our results confirm that ${\cal B}(B^0\rightarrow K^+\pi^-)$
is larger than ${\cal B}(B^0\rightarrow\pi^+\pi^-)$, and indicate that
${\cal B}(B^+\rightarrow h^+\pi^0)$ and ${\cal B}(B^0\rightarrow K^0\pi^0)$
seem to be larger than expected in relation to the $B^0\rightarrow h^+\pi^-$ 
and $B^+\rightarrow K^0\pi^+$ modes based on isospin or penguin
dominance arguments \cite{phys}.

We wish to thank the KEKB accelerator group for the excellent operation
of the KEKB accelerator.
We acknowledge support from the Ministry of Education, Culture, Sports,
Science, and Technology of Japan and
the Japan Society for the Promotion of Science;
the Australian Research Council and
the Australian Department of Industry, Science and Resources;
the Department of Science and Technology of India;
the BK21 program of the Ministry of Education of Korea and
the CHEP SRC program of the Korea Science and Engineering Foundation;
the Polish State Committee for Scientific Research under contract
No.2P03B 17017;
the Ministry of Science and Technology of Russian Federation;
the National Science Council and the Ministry of Education of Taiwan;
the Japan-Taiwan Cooperative Program of the Interchange Association;
and the U.S. Department of Energy.

\newpage

\newpage

\begin{table}[]
\caption{{}
	Summary of the results.
	The obtained signal yield ($N_s$), statistical significance
	($\Sigma$), efficiency ($\epsilon$), charge averaged
	branching fraction ($\cal B$) and its 90\% confidence level upper
	limit (U.L.) are shown.
	In the calculation of ${\cal B}$, the production rates
	of $B^+B^-$ and $B^0\overline{B}{}^0$ pairs are assumed to be equal.
	In the modes with $K^0$ mesons, $N_s$ and $\epsilon$ are quoted
	for $K^0_S$, while ${\cal B}$ and U.L. are for $K^0$.
	Submode branching fractions for $K^0_S\rightarrow\pi^+\pi^-$
	and $\pi^0\rightarrow\gamma\gamma$ are included in $\epsilon$.
	The first and second errors in $N_s$ and ${\cal B}$ are statistical
	and systematic errors, respectively.
}
\label{tab:table1}
\vspace{0.5cm}
\begin{tabular}{cccccc}

Mode & $N_s$ & $\Sigma$ &  $\epsilon$ [\%] &
${\cal B}$ [$\times 10^{-5}$] & U.L. [$\times 10^{-5}$]\\ \hline

$B^0\rightarrow\pi^+\pi^-$ & $17.7^{\ +7.1\ +0.3}_{\ -6.4\ -1.1}$ & 3.1 &
28.1 & $0.56^{\ +0.23}_{\ -0.20}\pm 0.04$ & -- \\

$B^+\rightarrow\pi^+\pi^0$ & $10.4^{\ +5.1\ +1.2}_{\ -4.3\ -1.6}$ & 2.7 &
12.0 & $0.78^{\ +0.38\ +0.08}_{\ -0.32\ -0.12}$ & $1.34$ \\

$B^0\rightarrow K^+\pi^-$ & $60.3^{\ +10.6\ +2.7}_{\ -9.9\ -1.1}$ & 7.8 &
28.0 & $1.93^{\ +0.34\ +0.15}_{\ -0.32\ -0.06}$ & -- \\

$B^+\rightarrow K^+\pi^0$ & $34.9^{\ +7.6\ +0.6}_{\ -7.0\ -2.0}$ & 7.2 &
19.2 & $1.63^{\ +0.35\ +0.16}_{\ -0.33\ -0.18}$ & -- \\

$B^+\rightarrow K^0\pi^+$ & $10.3^{\ +4.3\ +0.4}_{\ -3.6\ -0.1}$ & 3.5 &
13.5 & $1.37^{\ +0.57\ +0.19}_{\ -0.48\ -0.18}$ & -- \\

$B^0\rightarrow K^0\pi^0$ & $8.4^{\ +3.8\ +0.4}_{\ -3.1\ -0.6}$ & 3.9 &
9.4 & $1.60^{\ +0.72\ +0.25}_{\ -0.59\ -0.27}$ & -- \\

$B^0\rightarrow K^+K^-$ & $0.2^{\ +3.8}_{\ -0.2}$ & -- & 24.0 & --& 0.27 \\

$B^+\rightarrow K^+\overline{K}{}^0$ & $0.0^{\ +0.9}_{\ -0.0}$ & -- & 12.1 &
-- & 0.50 \\
\end{tabular}
\end{table}

\vspace{1.0cm}

\begin{table}[]
\caption{{}
	Ratio of charge averaged branching fractions (${\cal B}$) for
	$B\rightarrow\pi\pi$, and $ K\pi$ decays.
	The first error is statistical and the second is systematic.
	The correlation and cancellation of systematic errors are taken
	into account.
	A 90\% confidence level upper limit in the first ratio, is
	calculated using a similar method as the upper limit of ${\cal B}$
	described in text.
}
\label{tab:table2}

\vspace{0.5cm}
\begin{tabular}{l l}

Modes & Ratio \\ \hline
${\cal B}(B^+\rightarrow \pi^+\pi^0) /
 {\cal B}(B^0\rightarrow \pi^+\pi^-)$ &
$ < 2.67 $ \\

$2\,{\cal B}(B^+\rightarrow K^+\pi^0) /
 {\cal B}(B^0\rightarrow K^+\pi^-)$ & 
$ 1.69^{\ +0.46\ +0.17}_{\ -0.45\ -0.19} $ \\

${\cal B}(B^0\rightarrow\pi^+\pi^-) /
 {\cal B}(B^0\rightarrow K^+\pi^-)$ & 
$ 0.29^{\ +0.13\ +0.01}_{\ -0.12\ -0.02} $ \\

${\cal B}(B^0\rightarrow K^+\pi^-) /
 2\,{\cal B}(B^0\rightarrow K^0\pi^0)$ & 
$ 0.60^{\ +0.25\ +0.11}_{\ -0.29\ -0.16} $ \\

$2\,{\cal B}(B^+\rightarrow K^+\pi^0) /
 {\cal B}(B^+\rightarrow K^0\pi^+)$ &
$ 2.38^{\ +0.98\ +0.39}_{\ -1.10\ -0.26} $ \\

${\cal B}(B^0\rightarrow K^+\pi^-) /
 {\cal B}(B^+\rightarrow K^0\pi^+)$ &
$ 1.41^{\ +0.55\ +0.22}_{\ -0.63\ -0.20}$ 

\end{tabular}
\end{table}

\end{document}